Baseline Performance of AI Tools in Classifying Cognitive Demand of Mathematical Tasks

Danielle S. Fox[1], Brenda L. Robles[2], Elizabeth DiPietro Brovey[2], Christian D. Schunn[1,2]

[1]Learning Research and Development Center, University of Pittsburgh

[2]Institute for Learning, University of Pittsburgh

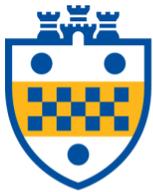
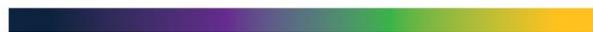

# Abstract

Teachers face increasing demands on their time, particularly in adapting high-quality mathematics curricula to meet individual student needs while maintaining cognitive rigor. This study evaluates whether artificial intelligence tools can accurately classify the cognitive demand of mathematical tasks, which we consider a foundational ability for creating or adapting tasks that support student learning. We tested eleven AI tools: six general-purpose tools (ChatGPT, Claude, DeepSeek, Gemini, Grok, Perplexity) and five education-specific tools (Brisk, Coteach AI, Khanmigo, Magic School, School.AI), on their ability to categorize mathematics tasks across four levels of cognitive demand using a widely-used, research-based framework for defining cognitive demand. The goal was not to determine whether prompts can be optimized to achieve better performance, but rather to approximate the performance teachers will achieve with a relatively straightforward prompt. On average, AI tools accurately classified task cognitive demand in only 63% of the cases. Education-specific tools overall were not more accurate than general-purpose tools, and no tool exceeded 83% accuracy. Notably, all tools struggled with tasks at the extremes of cognitive demand (*Memorization* and *Doing Mathematics*), exhibiting a systematic bias toward middle-category classifications (*Procedures with/without Connections*). The tools often gave plausible-sounding explanations for their decisions, which are likely to be persuasive to novice teachers. Detailed error analysis of instances in which AI tools misclassified the broad level of cognitive demand (high vs. low) revealed that the AI tools consistently over-weighted surface textual features rather than analyzing underlying cognitive processes. Further, unpacking AI tools' ability to reason about the factors that make a task higher vs. lower cognitive demand showed weaknesses across the board. That is, errors were made not by ignoring relevant dimensions of the task, but rather by incorrectly reasoning about multiple aspects of the task. These findings have important implications for the integration of AI into teacher planning workflows and highlight the need for improved prompt engineering and tool development for educational applications.

# 1. Introduction

## 1.1 Context and Motivation

Teachers are increasingly stretched thin, managing multiple responsibilities that extend far beyond classroom instruction. Even when provided with high-quality curricula, teachers must adapt materials to meet individual student needs, accommodate time constraints, and maintain pedagogical rigor.

The cognitive demand of mathematical tasks has been shown to significantly influence student learning: Tasks requiring deeper conceptual understanding and mathematical reasoning promote more robust learning than those focusing on memorization or procedural application without connections to underlying concepts (Stein & Lane, 1996; Stein et al., 1996). It is critical that teachers are supported in planning lessons that regularly involve high-quality tasks, regardless of the adaptations that are made for their students. However, identifying and modifying tasks to achieve appropriate cognitive demand requires substantial pedagogical expertise and time, with time being an increasingly scarce resource for in-service teachers.

Recent advances in artificial intelligence, particularly large language models (LLMs), have generated interest in their potential to reduce teacher workload. While existing research has primarily focused on student-facing AI applications such as intelligent tutoring systems (Hwang & Tu, 2021) and perspectives of pre-service teachers (e.g., Montero-Izquierdo et al., 2025), there is growing recognition that AI tools might support teachers in the process of planning and curriculum adaptation. Indeed, a number of such tools have been created. However, we know surprisingly little about whether current AI tools can perform fundamental pedagogical tasks such as accurately assessing the cognitive demand of instructional materials.

## 1.2 Literature Review and Research Gaps

The emerging literature on AI in education reveals several critical gaps. First, most studies focus on pre-service teachers experimenting with AI in artificial settings rather than experienced teachers integrating AI into authentic planning workflows (Aqazade et al., 2025; Kim et al., 2025; Lee & Yeo, 2022; Lee et al., 2025; Pelton & Pelton, 2024; Son et al., 2024; Uygun et al., 2025; Wijaya et al., 2025 ). Second, research has overwhelmingly focused on ChatGPT (e.g., Fock et al., 2025; Demir & Çimen, 2024; Lestari et al., 2025 Lee & Zhai, 2024; Küchemann et al., 2023; Pelton & Pelton, 2023; Sapkota & Bondurant, 2024; Şimşek , 2025; Tunç, 2024), leaving other AI tools largely unstudied despite their increased use in educational contexts. Third, there is insufficient subject-specific depth in existing research (Awang et al., 2025; Hwang & Tu, 2021), particularly in mathematics, an important subject for success in future STEM (Science, Technology, Engineering, Mathematics) courses and careers (Kohen et al., 2022; Maass et al., 2019).

Perhaps most critically, there is no systematic evidence regarding AI tools' ability to perform core pedagogical tasks such as evaluating cognitive demand. If AI tools cannot accurately assess the cognitive level of existing tasks, their utility for more complex planning activities like modifying tasks or creating new ones is undermined.

### 1.3 Theoretical Framework: The Task Analysis Guide

The present study employs the Task Analysis Guide (TAG) developed by Stein and Smith (1998) and is widely used as the basis of mathematics education research and mathematics teacher education/professional learning (See Appendix B for complete TAG). The TAG emerged from the [QUASAR](QUASAR) project, a five-year study of mathematics reform in urban middle schools, through extensive analysis of hundreds of classroom tasks and their relationship to student learning outcomes (Silver & Stein, 1996; Stein et al, 1996). A key finding from this research demonstrated that selecting high-level tasks is necessary for student engagement in complex thinking, whereas low-level tasks almost never result in high-level engagement. Thus, task selection is a critical starting point for instruction (Stein & Lane, 1996).

The TAG categorizes mathematical tasks into four levels of cognitive demand:

**Low Cognitive Demand:**

1. Memorization: Tasks that require reproduction of previously learned facts, rules, formulas, or definitions without connections to concepts or meaning

2. Procedures without Connections: Tasks that focus on producing correct answers using procedures without developing conceptual understanding

**High Cognitive Demand:**

3. Procedures with Connections: Tasks that focus students' attention on the use of procedures for the purpose of developing deeper understanding of mathematical concepts and ideas

4. Doing Mathematics: Tasks that require complex, non-algorithmic thinking and demand self-monitoring of one's own cognitive processes

The TAG provides specific characteristics for each category, offering a rigorous framework for task analysis that has been validated through extensive use in professional development and research contexts (Stein & Lane, 1996; Stein et al., 1996).

### 1.4 Research Questions

The present study addresses the following research questions:
1. Can current AI tools accurately classify the cognitive demand of mathematics tasks using an established framework (the Task Analysis Guide)?

2. How does classification accuracy vary across different AI tools (general-purpose vs. education-specific)?
3. Are certain types of tasks systematically more difficult for AI tools to classify accurately?
4. What patterns of misclassification emerge, and what do they reveal about AI tools' ability to evaluate cognitive demand?

## 2. Methods and Results

### 2.1 Task Selection and Expert Classification

*2.1.1 Methods*

We selected twelve mathematics tasks representing the four cognitive demand categories, with three exemplar tasks per category. All tasks had been previously analyzed and categorized by human experts trained in using the Task Analysis Guide. The tasks are representative of various mathematical content areas and grade levels and are regularly used in teacher training contexts to challenge teachers' thinking about the cognitive demand of tasks.

Tasks were labeled A through L and distributed across cognitive demand categories as follows:

**Memorization**: Tasks A, E, L

> Example: Task A
>
> *What are the decimal and percent equivalents for the fractions ½ & ¼?*

**Procedures without Connections**: Tasks D, H, J

> Example: Task D
>
> *Convert the fraction 3/8 to a decimal and a percent. Show your work.*

**Procedures with Connections**: Tasks B, G, I

> Example: Task B
>
> *Using a 10 x 10 grid, identify the decimal and percent equivalents of 3/8.*

**Doing Mathematics**: Tasks C, F, K

> Example: Task F

*Alazar Electric company sells light bulbs to big box stores – the big chain stores that frequently buy large numbers of bulbs in one sale. They sample their bulbs for defects.*

*A sample of 96 light bulbs consisted of 4 defective ones. Assume today's batch of 6,000 light bulbs has the same proportion of defective bulbs as the sample.*

*What are the total number of defective bulbs made today?*

*The big businesses agree to accept no larger than a 4% rate of defective bulbs.*

*Does today's batch meet that expectation? Explain in writing and with equations how you made your decision.*

All tasks can be found in Appendix C.

*2.1.2 Results: Performance by Task*

Classification accuracy varied dramatically across tasks ($M$ = 62%, range = 9 –100%; See Table 1).

**Table 1.** *Task classification accuracy across AI tools*

| Task | Correct Classification Category | Accuracy Across AI Tools |
|---|---|---|
| D | 2: Procedures without Connections | 100% |
| H | 2: Procedures without Connections | 100% |
| J | 2: Procedures without Connections | 100% |
| B | 3: Procedures with Connections | 91% |
| I | 4: Procedures with Connections | 82% |
| E | 1: Memorization | 55% |
| G | 3: Procedures with Connections | 55% |
| L | 1: Memorization | 45% |
| F | 4: Doing Mathematics | 45% |
| A | 1: Memorization | 33% |
| C | 4: Doing Mathematics | 27% |
| K | 4: Doing Mathematics | 9% |

A clear pattern emerged: tasks in the middle categories (*Procedures with/without Connections*) were classified with higher accuracy than tasks at the extremes (*Memorization* and *Doing Mathematics*).

## 2.2 AI Tool Selection

*2.2.1 Methods*

We selected eleven AI tools through a systematic process. A Google search identified commonly used AI platforms in the education sector, as well as the most widely used general-purpose AI tools.

**General-Purpose AI Tools (n = 6):**
- ChatGPT Free GPT 5
- Claude Free
- DeepSeek DeepSeek-V3.2-Exp
- Gemini Free 2.5 Flash
- Grok Free Version 4
- Perplexity Pro

**Education-Specific AI Tools (n = 5):**
- Brisk: Beta
- Coteach AI: Free
- Khanmigo Free
- Magic School: Free Plan
- School.AI: Free Plan

*2.2.2 Results: Performance by AI Tool*

Across all tasks, AI tools achieved an average classification accuracy of 62%, performing well above chance (25%) but far below expert-level performance. However, there was considerable variation across tools (See Table 2. Complete classification data can be found in Appendix D).

**Table 2.** *Average classification accuracy by AI tool*

| AI Tool (base tool) | Average Classification Accuracy |
|---|---|
| **General-Purpose Tools** | |
|     DeepSeek | 83% |
|     Grok | 67% |
|     Gemini | 58% |
|     Perplexity (Multi-modal) | 58% |
|     ChatGPT | 50% |
|     Claude | 50% |
| **Education-Specific Tools** | |
|     Coteach (Anthropic Claude) | 75% |
|     Magic School (Multi-modal) | 67% |
|     Brisk (Google Gemini) | 67% |
|     School AI (OpenAI ChatGPT) | 58% |
|     Khanmigo (Microsoft) | 50% |

Education-specific tools did not significantly outperform general-purpose tools (63% accuracy vs 61%, respectively). Coteach achieved the highest accuracy among specialized tools (75%), but DeepSeek, a general-purpose tool, achieved the highest overall accuracy (83%).

## 2.3 Prompting Procedure

*2.3.1 Methods*

For each of the twelve task classifications, we uploaded both the Task Analysis Guide and a single mathematics task in either .docx or .pdf format. We used the same standardized prompt across all tools and tasks:

> "Based on the 'TAG.docx' I uploaded, determine whether the math task 'Task[X].docx' uploaded is high-cognitive demand or low-cognitive demand. Give detailed reasoning and output the detailed analysis with dimensions in a table."

This "out of the box" approach intentionally avoided any prompt engineering techniques, establishing a baseline performance benchmark for how these tools function without specialized optimization.

*2.3.2 Results: Classification Distribution Patterns*

Analysis of classification choices revealed a systematic bias toward middle categories:

**Total Classifications by Category (133 total classifications):**
- Memorization: 15 (11%)
- Procedures without Connections: 53 (40%)
- Procedures with Connections: 50 (37%)
- Doing Mathematics: 15 (11%)

Despite equal representation of categories in the task set (3 tasks from each cognitive demand category), AI tools overwhelmingly classified tasks as *Procedures with* or *without Connections*. This suggests a central-tendency bias where AI tools default to middle categories when uncertain.

**Individual tool patterns varied:**
- ChatGPT never classified any task as *Memorization* or *Doing Mathematics*
- Gemini and School AI also avoided the highest category (*Doing Mathematics*)
- Only DeepSeek showed relatively balanced distribution across all four categories

A comprehensive table of findings can be found in Appendix D.

## 2.4 Error Analysis

*2.4.1 Methods*

For the present study, we conducted two types of error analyses. First, we identified instances where AI tools misclassified the broad level of cognitive demand (labeling a high-demand task as low-demand, or vice versa). These represent the most serious errors, as the tools fundamentally misinterpreted the nature of the math task. Second, for all broad-level misclassifications, we analyzed the reasoning provided by AI tools. Human experts examined each detailed output and deconstructed the AI's reasoning line by line to identify language that matched (exactly or semantically) the characteristics outlined in the Task Analysis Guide. For each characteristic from the correct classification category, experts coded the AI's reasoning as follows:

- **0: Not addressed** - The characteristic was not mentioned or considered
- **1: Addressed but incorrect** - The characteristic was discussed but with faulty reasoning
- **2: Addressed but only partially correct** - Some aspects were correct but incomplete
- **3: Addressed and correct** - The characteristic was fully and accurately addressed

For example, Task E, a *Memorization* task, asks students "True or false? 4/100 < 4/96." This was misclassified by Coteach as a *Procedures with Connections* task. In its output, it stated "[This task] cannot be solved through memorization or routine algorithms. Not a commonly memorized comparison."

Experts paired this reasoning with the first characteristic of the correct classification for *Memorization* which states, "[The task] involves reproducing previously learned facts, rules, formulae, or definitions." Experts then coded the sentences in the statement regarding this characteristic as incorrect (red) or correct (green) and gave the statement a score between 0 and 3. For Coteach's statement about Task E, a score of 2 was ascribed (Addressed but only partially correct) because this task *can* be solved through routine algorithms, but it is *not* a commonly memorized comparison.

This allowed us to systematically assess which cognitive demand characteristics AI tools misinterpreted or ignored and the reasoning they used to support their choice. Complete examples of AI responses can be found in Appendix E.

*2.4.2 Results: Broad-Level Misclassifications*

Our error analysis focused on broad-level errors where the AI failed to identify the correct overall cognitive demand level. For instance, when experts labeled a task as 'high demand', but the AI classified it as 'low demand'.

Of the 133 total classifications, there were only six (5%) instances of a tool misclassifying the broad level of cognitive demand. While this suggests AI tools generally distinguish between high and low demand, these errors are significant as they represent fundamental misinterpretations of some cognitive demand characteristics.

## Broad-Level Misclassifications:
- **Task E** (Claude and Coteach classified this *Memorization* task as high demand: 2 instances)
- **Task B** (Gemini classified this *Procedures with Connections* task as low demand: 1 instance)
- **Task G** (Claude, DeepSeek, and Khanmigo classified this *Procedures with Connections* task as low demand: 3 instances)

*2.4.3 Results: Patterns Across All Broad-Level Errors*

Several patterns emerge from the detailed coding analysis:

1. **Surface feature bias**: AI tools consistently over-weighted surface textual features (e.g., explicit mention of "algorithm" or "procedure") rather than analyzing the cognitive processes required.
2. **Incomplete characteristic analysis**: Even when AI tools addressed relevant TAG characteristics, they rarely scored a 3 (fully correct). Most scores ranged from 1.0 to 2.0, indicating systematic difficulties in applying the framework correctly.
3. **Task G as a critical failure point**: With three misclassifications representing half of all broad-level errors, Task G appears to contain features that particularly confound AI reasoning about cognitive demand.

These detailed findings underscore that even when AI tools achieve reasonable overall accuracy, their reasoning about cognitive demand characteristics remains flawed, with significant implications for educational applications.

*2.4.4 Results: Misclassifications Within Demand Levels*

Even when AI tools correctly identified the broader cognitive demand level, they often misclassified the specific category. The most common patterns were:

**Memorization Tasks (A, E, L):**
- Most often classified as *Procedures without Connections*
- Only correctly identified 44% of the time on average
- Tools appeared to see any use of a procedure as disqualifying memorization

**Doing Mathematics Tasks (C, F, K):**
- Most often classified as *Procedures with Connections*
- Only correctly identified 27% of the time on average
- Task K was particularly problematic (9% accuracy), with most tools defaulting to *Procedures with Connections*

**Procedures with Connections Tasks (B, G, I):**
- Generally well-classified (75% average accuracy)
- Task G was the exception, frequently downgraded to *Procedures without Connections*

**Procedures without Connections Tasks (D, H, J):**
- Universally well-classified (100% accuracy for all three tasks)
- Suggests these tasks have the clearest distinguishing features

# 3. Discussion

## 3.1 Interpretation of Findings

The 62% average accuracy represents a mixed picture for AI tool capabilities in classifying cognitive demand. On the one hand, this performance substantially exceeds chance performance and demonstrates that AI tools can engage with aspects of complex pedagogical frameworks such as the Task Analysis Guide. On the other hand, 62% accuracy falls far short of what would be needed for reliable use in educational planning. Even the top-performing AI tools made major errors.

The wide variation in task-level accuracy (9% to 100%) is quite informative. This variation suggests that cognitive demand classification is not uniformly difficult for AI; rather, certain task characteristics make classification easier or harder. The near-perfect performance on *Procedures without Connections* tasks and the poor performance on *Doing Mathematics* tasks (especially Task K) point to systematic differences in how AI tools process task features, and a need for more objective, boundary-defining language in the rubric used to train the tools.

## 3.2 The Middle-Category Bias

One of the most interesting findings is the overwhelming tendency of AI tools to classify tasks as *Procedures with or without Connections.* This "middle bias" appears across nearly all tools and suggests several possible explanations:

Frequency in training data: AI models may have encountered more examples of procedural tasks in their training data, leading to a prior bias toward these categories.

Feature ambiguity: The extreme categories may require more nuanced understanding. *Memorization* tasks must be distinguished from simple procedures, while *Doing Mathematics* tasks must demonstrate non-algorithmic complexity. Middle categories may have more obvious textual markers.

Central tendency behavior: When uncertain, AI systems may default to middle options as a form of computational hedging, like how humans tend to avoid extreme choices (e.g., 0 or 5 on a Likert scale) under uncertainty (Douven & Schupbach, 2017; Stevens, 1971).

Definitional overlap: The Task Analysis Guide distinguishes *Procedures with* and *without Connections* based on whether procedures develop conceptual understanding, which is a subtle distinction that may be difficult to infer from task text alone.

## 3.3 What Makes Tasks Easy or Hard to Classify?

The perfect accuracy on Tasks D, H, and J versus poor accuracy on Tasks C and K suggests that surface features matter significantly. Tasks D, H, and J contain minimal text ($M = 17$ words), and may contain clear language that distinguishes procedural work without conceptual depth. For instance, Task H clearly states, "Use cross products to solve the problem," which explicitly tells students the procedure needed to be successful. Whereas tasks C and K (both *Doing Mathematics* tasks) are wordier ($M = 63$ words) and may be harder to identify because they require recognizing the absence of well-known algorithmic paths rather than the presence of specific features.

Task K had the lowest accuracy rate at 9%, with only one tool (DeepSeek) correctly identifying it as *Doing Mathematics* while all others classified it as *Procedures with Connections*. The tools' reasoning centered on a few recurring themes: they saw the solution path as predictable and flagged the absence of open-ended exploration. However, experts classify Task K as *Doing Mathematics* because it situates mathematics in a real-world context, requires students to construct and explain their own equation; ultimately demanding a high level of cognitive engagement. The gap between expert and AI judgement here likely reflects the tools' tendency to underweight the explanation and modeling demands of the task and instead focus on the transparent cost structure described in the problem.

### 3.4 Practical Implications

These findings have several important implications for educational practice:

**Limited readiness for autonomous use:** With 62% accuracy, current AI tools are not yet reliable enough to independently classify task cognitive demand without human oversight or significant prompt engineering. Even tools designed for supporting teachers in mathematics education (e.g., Coteach) need further guidance.

**Potential for decision support:** However, AI tools might serve effectively as decision-support systems, flagging tasks for human review or providing preliminary classifications that teachers verify.

**Need for prompt engineering:** The baseline "out of the box" approach likely underestimates potential performance. Optimized rubrics, prompts, exemplars, and iterative refinement could substantially improve accuracy (Schulhoff et al., 2024). Research-backed refinements are already underway, testing whether more structured prompts and exemplars can meaningfully improve alignment with expert classifications (Fox et al., in prep).

**Tool selection matters:** The range from 50% to 83% accuracy suggests that selecting the right tool for this task is consequential.

### 3.5 Future Directions

This baseline study opens several avenues for future research that would address current limitations and extend our understanding of AI tools' capacity to assess cognitive demand.

**Expanding task samples and contexts:** While our twelve tasks were carefully selected to represent each cognitive demand category and are regularly used in teacher training, they represent a limited sample of the full range of mathematical tasks teachers encounter. Future research should test AI classification accuracy across a larger, more diverse set of tasks spanning different mathematical content areas, grade levels, and task formats. Additionally, measuring inter-rater reliability among human experts would help distinguish between AI errors and genuinely ambiguous tasks that challenge even experienced educators.

**Optimizing AI performance through prompt engineering:** Our deliberate choice to use an "out of the box" prompt establishes a baseline but does not represent optimal performance potential. Systematic investigation of prompting strategies could substantially improve accuracy. Few-shot learning approaches that provide AI tools with example classifications before asking them to classify new tasks may support AI tools' calibration to the Task

Analysis Guide framework. Similarly, [chain-of-thought prompting](#) that requires step-by-step reasoning might help AI tools better distinguish subtle category differences, particularly between *Procedures with* and *without Connections*. Research comparing different prompting approaches would provide practical guidance for educators seeking to use these tools effectively. Though learning [prompt engineering](#) techniques is yet another time-consuming task for which teachers may not have time.

**Enabling iterative refinement:** Our study captured single-pass classifications without allowing AI tools to revise their assessments based on feedback. However, in authentic planning contexts, teachers might engage in dialogue with AI tools, providing corrective feedback that improves subsequent classifications. Future research should investigate whether and how AI tools can learn from corrections within a conversation, and whether this iterative approach narrows the performance gap with human experts.

**Multi-tool ensemble approaches:** Given the variation in accuracy across tools (50% to 83%), combining predictions from multiple AI tools might achieve higher reliability than any single tool. Ensemble methods could identify consensus classifications or flag tasks where tools disagree for human review. Research on optimal ensemble strategies and decision rules would be valuable for practical implementation.

**Contextualizing AI performance with human baselines:** Our findings show that AI tools achieve approximately 60% accuracy, but we lack systematic comparison data on how teachers with varying levels of training perform on these same tasks. Assessing classification accuracy among pre-service teachers, early-career teachers, and experienced teachers who have received professional development on the Task Analysis Guide would contextualize AI performance and help determine whether AI tools might serve as useful training aids or decision-support systems.

**Tracking temporal changes:** AI models update frequently, and our findings represent performance at a specific point in time. Longitudinal studies that periodically reassess the same tools on the same tasks would reveal whether classification accuracy improves as models evolve, and whether newer models address the systematic biases we observed (e.g., [central tendency bias,](#) surface feature over-weighting).

**Investigating reasoning quality beyond accuracy:** Our detailed error analysis revealed that even when AI tools achieved correct classifications, their reasoning often scored between 1.0 and 2.0 on our coding scale, indicating incomplete or partially incorrect application of Task Analysis Guide characteristics. Future research should examine whether improving reasoning quality, even when final classifications are correct, matters for educational applications, particularly if teachers rely on AI-generated explanations to deepen their own understanding of cognitive demand.

## 4. Conclusion

This study provides the first systematic evaluation of the ability of general-purpose and specialized AI tools to classify the cognitive demand of mathematics tasks using an established pedagogical framework. While AI tools demonstrate above-chance performance, their 62% average accuracy and strong bias toward middle categories reveal important limitations. The dramatic variation in accuracy across tasks (from 9% to 100%) suggests that task characteristics significantly influence AI performance, with extreme categories (*Memorization* and *Doing Mathematics*) proving particularly challenging.

These findings suggest that current AI tools are not yet ready for autonomous deployment in assessing cognitive demand but show promise as decision-support systems that could augment teacher expertise. The variation across tools indicates that thoughtful tool selection matters, and the baseline performance documented here establishes a foundation for future work on optimization strategies such as [prompt engineering](#).

As AI tools become increasingly integrated into educational contexts, understanding their capabilities and limitations for core pedagogical tasks is essential. This research contributes to that understanding while highlighting the continued irreplaceable value of teacher expertise in curriculum adaptation and instructional planning.

**Acknowledgements**

The authors wish to thank Mohammed Ghassemi and Ghamut for their consultation and expertise regarding artificial intelligence. Their insights and support contributed meaningfully to the development of this work. Additionally, the authors used Claude (Anthropic) to assist with organizing and structuring this manuscript. All intellectual content, analyses, and conclusions are solely the authors' own.

## Appendix A: Glossary of Terms

**Central tendency bias**: the tendency for raters to avoid extreme classifications and instead gravitate toward middle categories, resulting in an underrepresentation of the highest and lowest levels of a scale.

**Chain-of-thought prompting:** a prompting technique in which the AI tool is explicitly instructed to reason through a problem step by step before arriving at a final answer, encouraging more deliberate and transparent reasoning rather than jumping directly to a conclusion.

**Few-shot learning**: providing an AI model with a handful of labeled examples within the prompt to guide its responses on similar tasks.

**Out of the box**: the default, unmodified approach of querying AI tools without any additional prompt engineering, customized rubric language, or task-specific guidance — simply presenting the task as-is and accepting the tool's initial classification.

**Prompt engineering:** the practice of deliberately crafting and refining the inputs given to an AI tool, including instructions, context, and examples, in order to guide the model toward more accurate or consistent outputs.

**QUASAR** (Quantitative Understanding: Amplifying Student Achievement and Reasoning) project: an educational reform project aimed at fostering and studying the development and implementation of enhanced mathematics instructional programs for students attending middle school in economically disadvantaged communities (Silver & Stein, 1996).

# Appendix B: Mathematical Task Analysis Guide

| **Lower-Level Demands** | **Higher-Level Demands** |
|---|---|
| Memorization Tasks<br>• Involves either producing previously learned facts, rules, formulae, or definitions OR committing facts, rules, formulae, or definitions to memory.<br>• Cannot be solved using procedures because a procedure does not exist or because the time frame in which the task is being completed is too short to use a procedure.<br>• Are not ambiguous – such tasks involve exact reproduction of previously seen material and what is to be reproduced is clearly and directly stated.<br>• Have no connection to the concepts or meaning that underlie the facts, rules, formulae, or definitions being learned or reproduced. | Procedures with Connections Tasks<br>• Focus students' attention on the use of procedures for the purpose of developing deeper levels of understanding of mathematical concepts and ideas.<br>• Suggest pathways to follow (explicitly or implicitly) that are broad general procedures that have close connections to underlying conceptual ideas as opposed to narrow algorithms that are opaque with respect to underlying concepts.<br>• Usually are represented in multiple ways (e.g., visual diagrams, manipulatives, symbols, problem situations). Making connections among multiple representations helps to develop meaning.<br>• Require some degree of cognitive effort. Although general procedures may be followed, they cannot be followed mindlessly. Students need to engage with the conceptual ideas that underlie the procedures in order to successfully complete the task and develop understanding. |
| Procedures without Connections Tasks<br>• Are algorithmic. Use of the procedure is either specifically called for, or its use is evident based on prior instruction, experience, or placement of the task.<br>• Require limited cognitive demand for successful completion. There is little ambiguity about what needs to be done and how to do it.<br>• Have no connection to the concepts or meaning that underlie the procedure being used.<br>• Are focused on producing correct answers rather than developing mathematical understanding.<br>• Require no explanations, or explanations that focus solely on describing the procedure that was used. | Doing Mathematics Tasks<br>• Requires complex and non-algorithmic thinking (i.e., there is not a predictable, well-rehearsed approach or pathway explicitly suggested by the task, task instructions, or a worked-out example).<br>• Requires students to explore and to understand the nature of mathematical concepts, processes, or relationships.<br>• Demands self-monitoring or self-regulation of one's own cognitive processes.<br>• Requires students to access relevant knowledge and experiences and make appropriate use of them in working through the task.<br>• Requires students to analyze the task and actively examine task constraints that may limit possible solution strategies and solutions.<br>• Requires considerable cognitive effort and may involve some level of anxiety for the student due to the unpredictable nature of the solution process required. |

# Appendix C: Mathematics Tasks A-L

**Task A:**
What are the decimal and percent equivalents for the fractions ½ and ¼?

**Task B:**
Using a 10 x 10 grid, identify the decimal and percent equivalents of 3/5.

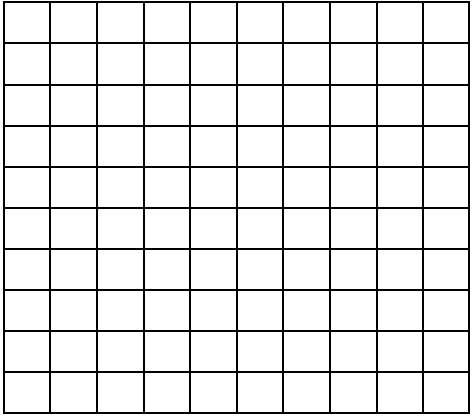

**Task C:**
Shade 6 of the small squares in the rectangle below.

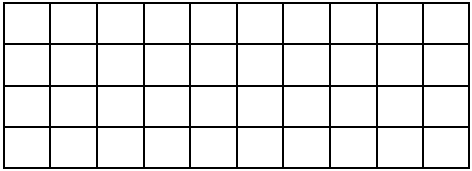

Using the diagram, explain how to determine each of the following:
- the percent of the area that is shaded
- the decimal part of the area that is shaded
- the fractional part of the area that is shaded

**Task D:**
Convert the fraction 3/8 to a decimal and a percent. Show your work.

**Task E:**
True or False? 4/100 < 4/96

**Task F:**

Alazar Electric company sells light bulbs to big box stores – the big chain stores that frequently buy large numbers of bulbs in one sale. They sample their bulbs for defects.

A sample of 96 light bulbs consisted of 4 defective ones. Assume today's batch of 6,000 light bulbs has the same proportion of defective bulbs as the sample.
- What are the total number of defective bulbs made today?

The big businesses agree to accept no larger than a 4% rate of defective bulbs.

Does today's batch meet that expectation? Explain in writing and with equations how you made your decision.

**Task G:**

Alazar Electric company sells light bulbs to big box stores – the big chain stores that frequently buy large numbers of bulbs in one sale. They sample their bulbs for defects.

A sample of 96 light bulbs consisted of 4 defective ones. Assume today's batch of 6,000 light bulbs has the same proportion of defective bulbs as the sample.
- Set up a proportion and solve it to determine the total number of defective bulbs.

The big businesses they sell to accept no larger than a 4% rate of defective bulbs.
Is today's batch less than 4% defective? Show your work.

**Task H:**

Use cross products to solve the proportion. Show your work.
4/96 = x/6000

**Task I:**

The table below shows the price in dollars that Custom T-Shirts charges a customer for a given number of t-shirts.

What equation can be used to determine how much to charge a customer for any number of shirts? Explain how you determined your answer.

| Number of Shirts | 1 | 2 | 3 | 4 | 5 |
|---|---|---|---|---|---|
| Price in Dollars | 23 | 31 | 39 | 47 | 55 |

**Task J:**
If y = 8x + 15, evaluate y when x =

        A.   10
        B.   20
        C.   30

Explain the procedure you used to find your solution.

**Task K:**
This past summer, you were hired to work at Custom T-Shirts. When a customer places an order for a special design, Custom T-Shirts charges a one-time fee of $15 to set up a t-shirt design, plus $8 for each t-shirt printed.

What equation can be used to determine how much to charge a customer for any number of shirts? Explain how you determined your answer.

**Task L:**
In the following equations, name the slope and y-intercept.
       A.   y = 8x + 15
       B.   y = -2x + 7
       C.   y = 4x - 9

# Appendix D: Complete Classification Data

**Table D1**

*AI Tool Classifications of Tasks A-L*

|  | Task A | Task B | Task C | Task D | Task E | Task F | Task G | Task H | Task I | Task J | Task K | Task L |
|---|---|---|---|---|---|---|---|---|---|---|---|---|
| **Correct Answer** | 1 | 3 | 4 | 2 | 1 | 4 | 3 | 2 | 3 | 2 | 4 | 1 |
| **Brisk** | 2 | 3 | 4 | 2 | 1 | 4 | 4 | 2 | 3 | 2 | 3 | 2 |
| **ChatGPT** | 2 | 3 | 3 | 2 | 2 | 3 | 3 | 2 | 3 | 2 | 3 | 2 |
| **Claude** | 1 | 3 | 3 | 2 | **4** | 3 | **2** | 2 | 3 | 2 | 3 | 2 |
| **Coteach** | 1 | 3 | 3 | 2 | **3** | 4 | 3 | 2 | 3 | 2 | 3 | 1 |
| **DeepSeek** | 1 | 3 | 4 | 2 | 1 | 4 | **2** | 2 | 4 | 2 | 4 | 1 |
| **Gemini** | 1 | **2** | 3 | 2 | 1 | 3 | 3 | 2 | 3 | 2 | 3 | 2 |
| **Grok** | 2 | 3 | 3 | 2 | 1 | 4 | 3 | 2 | 4 | 2 | 3 | 1 |
| **Khanmigo** | 2 | 3 | 3 | 2 | 2 | 3 | **2** | 2 | 3 | 2 | 3 | 1 |
| **Magic School** | 2 | 3 | 4 | 2 | 1 | 3 | 3 | 2 | 3 | 2 | 3 | 2 |
| **Perplexity** | 2 | 3 | 3 | 2 | 2 | 4 | 4 | 2 | 3 | 2 | 3 | 1 |
| **School. AI** | 2 | 3 | 3 | 2 | 1 | 3 | 3 | 2 | 3 | 2 | 3 | 2 |

*Note.* Abbreviations for categories of cognitive demand are 1 = Memorization, 2 = Procedures without Connections, 3 = Procedures with Connections, 4 = Doing Mathematics. Highlighted cells represent broad misclassifications.

# Appendix E: Detailed AI Response Examples

## E1: ChatGPT response for Task A

**TASK A**

Nice — I reviewed the TAG (Mathematical Task Analysis Guide) and the Task A file you uploaded, then evaluated the task against the TAG dimensions. (Note: when the Task A text was extracted the two fraction symbols did not render — the prompt reads "What are the decimal and percent equivalents for the fractions and ?" — but that missing rendering does **not** change the cognitive-demand analysis because the task structure is clear: convert given fractions to decimals and percents.)

**Conclusion**

**Classification:** Low-cognitive demand — *Procedures Without Connections*.
 **Short justification:** The task asks students to perform routine conversions (fraction → decimal → percent). It is algorithmic, has a clearly prescribed procedure, requires little to no conceptual explanation, and focuses on producing correct numeric answers rather than on developing deeper conceptual connections. This matches the TAG description of "Procedures Without Connections."

**Detailed analysis (dimensions)**

| Dimension | Evidence from Task A | TAG criterion / interpretation | Impact on cognitive demand |
|---|---|---|---|
| Task type (TAG categories) | Convert given fractions to decimal and percent. | Matches **Procedures Without Connections**: algorithmic conversion. TAG | Low demand |
| Clarity / ambiguity | Task statement is direct (asks for equivalents). (Note: fraction symbols did not render in extracted text, but the instruction form is unambiguous.) Task A | TAG: procedures tasks "little ambiguity about what needs to be done." TAG | Low demand |
| Procedural vs. conceptual | Requires use of a standard procedure (divide numerator by denominator → multiply by 100). No prompt to explain why or connect to fraction meaning. | TAG: procedures without connections "have no connection to the concepts or meaning… focused on producing correct answers." TAG | Low demand |
| Representation / multiple representations | Expected representations: numeric decimal and percent. Task does not require or suggest multiple representations (visual diagrams, models). | TAG: high-level tasks often require multiple representations; this task does not. | Low demand |
| Degree of cognitive effort / non-routine thinking | Routine arithmetic conversion; predictable sequence. | TAG: low cognitive demand when algorithmic and can be followed mindlessly. | Low demand |
| Requirement for explanation / justification | Task asks only for equivalents (answers), not for explanation of procedure or reasoning. | TAG: procedures without connections "require no explanations, or explanations that focus solely on describing the procedure." | Low demand |
| Potential for student sense-making or reasoning | Minimal unless teacher extends the task (e.g., ask why conversion works or connect to proportional reasoning). | TAG: to reach higher demand the task must prompt connections or non-algorithmic thinking. | Low demand as written |

| Multiple solution paths or problem-solving | Single straightforward method (division or known equivalent conversions). | TAG: "Doing Mathematics" requires non-algorithmic thinking and multiple pathways — absent here. | Low demand |

**Notes & suggestions (if you want to raise the cognitive demand)**

If your goal is to make Task A a **higher-demand** task, you could modify it in one of these ways (examples):

- Ask students to **explain** the relationship between fraction, decimal and percent representations (connect procedure to meaning). (Moves toward *Procedures with Connections*.)
- Provide an unfamiliar denominator and ask students to **justify** a method for quick conversion without a calculator (promotes reasoning).
- Give a contextual problem that requires selecting which form (fraction/decimal/percent) is most appropriate and justify the choice (moves toward *Doing Mathematics* if non-routine reasoning is required).
  All of these changes align with TAG's higher-level criteria.

If you'd like, I can (A) produce a revised version of Task A that pushes it to **Procedures with Connections** or **Doing Mathematics**, or (B) show a model solution for the current task (once you confirm the exact fractions, since the extraction didn't show the fraction numerators/denominators). Which would you prefer?

# E2: Claude's analysis of Task G

**Analysis**

**Task G is a LOW-COGNITIVE DEMAND task**, specifically classified as a **Procedures Without Connections Task**.

**Detailed Reasoning**

While Task G has the same contextual scenario as Task F, the specific task instructions fundamentally change the cognitive demand. This falls under low-cognitive demand because:

1. **Explicitly Prescribed Algorithm**: The phrase "Set up a proportion and solve it" removes student decision-making. The procedure is:
    a. Directly stated: students must use a proportion
    b. No choice in approach or strategy
    c. The exact method is prescribed, not suggested or implied
2. **Algorithmic Nature**: Students simply:
    a. Write the proportion: 4/96 = x/6,000
    b. Cross-multiply: 96x = 24,000
    c. Solve: x = 250
    d. This is a mechanical, step-by-step procedure
3. **Limited Cognitive Demand**: There is minimal ambiguity:
    a. What to do: set up a proportion (explicitly stated)
    b. How to do it: standard proportion-solving algorithm (well-rehearsed)
    c. No need to analyze which method to use
4. **"Show Your Work" ≠ Explanation**:
    a. "Show your work" means display calculation steps
    b. Unlike Task F's "Explain in writing and with equations how you made your decision"
    c. No requirement to explain reasoning, justify method choice, or demonstrate understanding
    d. Focus is on correct execution of the prescribed procedure
5. **No Conceptual Connection Required**: Students can successfully complete this by:
    a. Mechanically following the proportion algorithm
    b. Without understanding why proportions work
    c. Without connecting to the real-world context meaningfully
    d. The procedure is opaque with respect to underlying concepts
6. **Focus on Correct Answers**: The task prioritizes:
    a. Proper setup of the proportion
    b. Correct execution of solving procedures
    c. Right numerical answers
    d. NOT on developing understanding of proportional relationships
7. **Second Part Also Procedural**: "Is today's batch less than 4% defective?" becomes:
    a. Calculate 250/6,000 = 0.04166… = 4.17%
    b. Compare to 4%
    c. Answer: No
    d. This is straightforward computation and comparison without required justification

**Critical Difference from Task F**

**Task F** (HIGH cognitive demand):

- "Explain in writing and with equations how you made your decision"
- Students choose their approach
- Emphasis on understanding and justification
- Connections between procedures and concepts explicit

**Task G** (LOW cognitive demand):

- "Set up a proportion and solve it"
- Method is prescribed
- "Show your work" = show calculation steps
- Can be completed mechanically without understanding

**Comparison Table**

| Dimension | Task G Characteristics | Classification |
|---|---|---|
| **Cognitive Process** | Mechanically applying prescribed proportion-solving algorithm | Procedures Without Connections |
| **Procedure Usage** | **Explicitly called for**: "Set up a proportion and solve it" | Low-level |
| **Ambiguity Level** | **None** - exactly what procedure to use and how is evident | Low-level |
| **Conceptual Connection** | **No connection required** - can solve mechanically without understanding why | Low-level |
| **Explanation Required** | "Show your work" = display steps, NOT explain reasoning or meaning | Low-level |
| **Student Choice** | **No choice** in method or approach - proportion is prescribed | Low-level |
| **Cognitive Effort** | **Minimal** - straightforward application of algorithm | Low-level |
| **Solution Pathway** | **Completely predictable** - exact procedure specified | Low-level |
| **Real-World Context** | Present but not meaningfully engaged with; just window dressing | Low-level |
| **Focus** | **Producing correct answers** through proper algorithm execution | Low-level |
| **Procedure Transparency** | **Opaque** - proportion algorithm doesn't reveal conceptual understanding | Low-level |
| **Mindless Execution?** | **Yes** - can be completed by following steps without understanding context or concepts | Low-level |
| **Decision-Making** | **None required** - method is told, comparison is straightforward | Low-level |
| **Justification** | **Not required** - just show calculations | Low-level |
| **Overall Classification** | **Procedures Without Connections Task** | **LOW-COGNITIVE DEMAND** |

**Why the Subtle Wording Changes Matter**

The difference between Task F and Task G demonstrates how small instructional changes dramatically affect cognitive demand:

| Task F (HIGH) | Task G (LOW) |
|---|---|
| "What are the total number…" (open) | "Set up a proportion and solve it" (prescribed) |
| Students choose method | Method is dictated |
| "Explain in writing and with equations how you made your decision" | "Show your work" |
| Requires justification and reasoning | Requires calculation display only |
| Connects procedures to concepts | Procedures disconnected from meaning |
| Develops understanding | Focuses on answer production |

**Comparison to All Previous Tasks**

- **Task A**: Memorization (LOW)
- **Task B**: Procedures with connections - visual representation (HIGH)
- **Task C**: Procedures with connections - explanation focused (HIGH)
- **Task D**: Procedures without connections - algorithmic conversion (LOW)
- **Task E**: Doing mathematics - conceptual comparison (HIGH)
- **Task F**: Procedures with connections - real-world with justification (HIGH)
- **Task G**: **Procedures without connections - prescribed algorithm (LOW)**

**Important Pedagogical Insight**

This comparison between Tasks F and G illustrates a critical point for teachers:

**The same problem context can be either high or low cognitive demand depending on how the task is framed and what is asked of students.**

- Adding "Explain how you made your decision" → HIGH cognitive demand
- Prescribing "Set up a proportion" → LOW cognitive demand

Task G demonstrates how well-intentioned real-world contexts don't automatically create high cognitive demand. The instructional framing and required student actions determine the actual cognitive level, not just the context or problem scenario.

# E3: Gemini's analysis of Task L

The math task "Task L" is a **low-cognitive demand** task[1].
It falls under the specific category of **Procedures Without Connections Tasks** [2], as it involves using a well-known algorithm (identifying the slope and y-intercept in the standard slope-intercept form $y=mx+b$) which is evident based on prior instruction and requires limited cognitive effort for successful completion[3].

 **Task L Analysis: Low-Cognitive Demand**

Task L asks students to "In the following equations, name the slope and y-intercept" for equations in the form $y=mx+b$[4]. This is a clear application of the procedure for reading the slope ($m$) and y-intercept ($b$) directly from the equation[5].

| Demand Dimension | Analysis for Task L | Supporting Criteria from TAG.docx | |
|---|---|---|---|
| **Cognitive Demand Level** | | **Lower-Level Demands** [6] | The task requires limited cognitive demand for successful completion[7]. |
| **Specific Task Type** | | **Procedures Without Connections Tasks** [8] | The task is **algorithmic**[9]. |
| **Procedure Use** | The use of the procedure (identifying $m$ and $b$ in $y=mx+b$) is **evident** and directly called for by the task[10][10]. | Use of the procedure is either specifically called for or its use is **evident based on prior instruction**[11]. | |
| **Ambiguity** | There is **little ambiguity**[12]. Students simply read two values from the equation. | There is little ambiguity about what needs to be done and how to do it[13]. | |
| **Conceptual Connection** | The task **has no connection** to the underlying concepts of slope (rate of change) or y-intercept (initial value)[14]. | Tasks have **no connection to the concepts or meaning** that underlie the procedure being used[15]. | |
| **Focus** | The focus is purely on producing the **correct answers** (the correct numbers for $m$ and $b$)[16]. | Tasks are focused on producing correct answers rather than developing mathematical understanding[17]. | |
| **Explanation Requirement** | The task **does not require explanations**[18]. If an explanation were required, it would solely describe the procedure. | Tasks require **no explanations**, or explanations that focus solely on describing the procedure that was used[19]. | |

Would you like me to analyze another math task based on the criteria in the "TAG.docx" file?